\documentclass[11pt]{article}
\usepackage[margin=1in]{geometry}
\usepackage{authblk}      
\usepackage{graphicx,amsmath,amssymb, amsfonts}
\usepackage[numbers,sort&compress]{natbib}
\usepackage{caption}
\usepackage[colorlinks=true,allcolors=blue]{hyperref}
\DeclareMathOperator{\erf}{erf}

\title{Evaluating the predictive power of pump-probe imaging contrast of melanin for metastatic outcome: A 71-patient study}

\author[a]{David Grass}
\author[b]{Emma Bao}
\author[a,b]{Martin C. Fischer}
\author[c]{M. Angelica Selim}
\author[d,e,f]{Georgia M. Beasley}
\author[a,b,f,g,h,*]{Warren S. Warren\footnote{warren.warren@duke.edu}}
\affil[a]{Duke University, Department of Chemistry, Durham, North Carolina, USA}
\affil[b]{Duke University, Department of Physics, Durham, North Carolina, USA }
\affil[c]{Duke University, Department of Pathology, Durham, North Carolina, USA}
\affil[d]{Duke University, Department of Medicine, Durham, North Carolina, USA}
\affil[e]{Duke University, Department of Surgery, Durham, North Carolina, USA}
\affil[f]{Duke University, Duke Cancer Institute, Durham, North Carolina, USA}
\affil[g]{Duke University, Department of Biomedical Engineering, Durham, North Carolina, USA}
\affil[h]{Duke University, Department of Radiology, Durham, North Carolina, USA}

\begin{document} 
\maketitle

\begin{abstract}
\textbf{Significance:} Most melanoma deaths arise from metastatic spread, yet current staging imperfectly identifies which primary tumors will progress. Melanin structure is altered during malignant transformation, and pump-probe microscopy (PPM) measures melanin excited-state dynamics in standard biopsy sections, offering molecular contrast that is complementary to morphology-based assessment.

\textbf{Aim:} To determine whether PPM-derived melanin excited-state dynamics in primary cutaneous melanoma can predict metastatic outcome.

\textbf{Approach:} We imaged unstained sections from 71 primary cutaneous melanomas and 18 melanoma metastases. Transient absorption curves were fit to a biexponential excited-state model, and the fit parameters were used as features for learning classifiers to predict metastatic outcome.

\textbf{Results:} Melanin dynamics separated primary from metastatic tissue: both excited-state lifetimes were shorter in primaries, with common-language effect sizes up to 0.26. The same parameters did not separate primary tumors by outcome (effect sizes 0.41–0.61), and univariate logistic regression found no significant association for either the excited-state lifetimes or the ESA/GSB ratio. The best classifier reached 74\% patient-level accuracy (AUC 0.73), a plateau reached by several unrelated architectures.

\textbf{Conclusions:} PPM robustly distinguishes primary from metastatic melanoma tissue, consistent with progressive melanin disaggregation. While it is clear that pump-probe features of melanin thus correlate with tissue differences, melanin dynamics in single sections of the primary tumor, confounded by heterogeneity, do not give robust predictions of metastatic outcome.
\end{abstract}




\section{Introduction}
Cutaneous melanoma is the malignancy of melanocytes, the pigment-producing cells of the epidermis. Although the 5-year survival rate of 94.7\% is favorable for most patients, outcomes for those with metastatic disease remain poor despite advances in immunotherapy and targeted therapy \cite{seer_melanoma_2026, boutros_treatment_2024, joshi_cutaneous_2025}. Because the majority of melanoma deaths arise from metastatic spread, early and accurate identification of metastatic potential from the primary biopsy remains a critical clinical objective.

Current staging and prognosis rely on visual and dermatoscopic inspection, histopathological examination of the primary tumor, and, when necessary, lymph node biopsy \cite{keung_eighth_2018}. Histopathological features remain the diagnostic gold standard, but they imperfectly identify the most aggressive tumors. Gene expression profiling assays and liquid biomarkers have entered clinical use but still require further development \cite{kaminska_liquid_2021, chan_circulating_2024, eggermont_identification_2020, Lim2018, Yousaf2021a, Marchetti2020}, and AI- and machine-learning-based approaches trained on dermatoscopic or histopathological images \cite{kulkarni_deep_2020, yan_deep_2025, lallas_deep_2024, lallas_deep_2025} rely, like pathologists, on morphology rather than molecular information.

Melanoma tumors are typically heavily pigmented because melanocytes proliferate rapidly and synthesize melanin, a complex, heterogeneous biopolymer whose structure is known to change during malignant transformation \cite{wakamatsu_2023}. Pump-probe microscopy (PPM) \cite{Fischer2016} can measure the excited-state dynamics of melanin and provide spatially resolved contrast sensitive to melanin species, aggregation state, and oxidation \cite{Fu2008, Simpson2013, simpsonNearInfraredExcitedState2014, Ju2018}. Prior work by our group and others has used PPM to distinguish nevi from melanoma, to characterize melanin disaggregation, and to show that pump-probe signatures correlate with metastatic potential \cite{Matthews2011, wilsonImagingMicroscopicPigment2013a, Robles2015, Robles2017, Ju2019, Grass2022b, zhouRecentAdvancesNonlinear2025}. These observations motivate the hypothesis that the aggressive tumor microenvironment disaggregates melanin from highly assembled protomolecules into smaller subunits, and that this change informs on metastatic potential \cite{simpsonNearInfraredExcitedState2014, Ju2018, Ju2019, Grass2022b}.

In this manuscript we measure transient absorption (TA) with PPM and use machine learning to ask whether the TA dynamics of melanin in primary cutaneous melanoma predict metastatic outcome. We image primary tumor biopsies from 71 patients together with metastatic biopsies from 18 patients. We find that melanin excited-state dynamics reliably separate \emph{primary} from \emph{metastatic tissue} at the cohort level: lifetimes are consistently shorter in primaries than in metastases, consistent with the disaggregation hypothesis. However, these same features do not distinguish primary tumors by their eventual metastatic \emph{outcome}. Univariate predictors proposed in earlier small-cohort work, the excited-state lifetime and the ESA-to-GSB ratio, show no significant association with outcome. We further find by benchmarking six machine learning classifiers under three data-aggregation strategies that prediction plateaus near 74\% patient-level accuracy across very different model architectures. We believe this ceiling is caused by intrinsic tissue heterogeneity and signal-to-noise limitations, and we close by discussing the promise and current limitations of the approach, including the likely need for volumetric imaging and larger cohorts to localize the metastatic signal within primary tumors.

\section{Material and Methods}
\label{ch:material_methods}
    \subsection{Pump-Probe Microscopy}
    Pump-probe microscopy is a combination of pump-probe spectroscopy and laser scanning microscopy \cite{Fischer2016}. A pump pulse excites electronic and vibrational populations in a sample and a probe pulse is used to monitor these populations relaxing back to thermal equilibrium. Different molecules exhibit different population dynamics, which are further affected by the molecules' environment. The molecular dynamics are apparent in the measured transient absorption (TA) curves, the change of probe absorption as a function of the pump-probe inter-pulse delay $\Delta t$. Many nonlinear interactions between laser pulses and sample can occur, such as multiphoton absorption, stimulated Raman scattering, and excited state absorption. The multiplicity of the physical interactions and variability of the excited state lifetimes are the origin of the high chemical specificity of pump-probe microscopes.
    \begin{figure}[h!]
    \centering
        \includegraphics{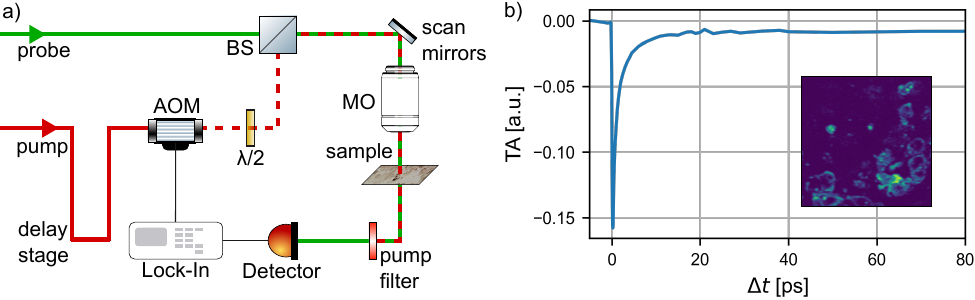}
        \caption{a) Pump-probe microscope experimental setup with beam splitter (BS), microscope objective (MO), and acousto-optical modulator (AOM). b) Spatially average transient absorption curve of primary melanoma tumor (inset is a image of the integrated signal).}
        \label{fig:experimental_setup}
    \end{figure}
    The experimental setup is similar to \cite{Grass2022b} and shown in Figure~\ref{fig:experimental_setup}a). Two time-synchronized ultrafast laser pulse trains are used as pump and probe pulses. The color of both laser sources are tunable and throughout this manuscript two wavelength combinations are used: either a pump wavelength of $\lambda_\text{pump}=770$nm and probe wavelength of $\lambda_\text{probe}=730$nm (labeled as 770/730) or a pump wavelength of $\lambda_\text{pump}=730$nm and probe wavelength of $\lambda_\text{probe}=810$nm (labeled as 730/810) is used. Both pulse trains have a repetition rate of 80MHz and their pulse widths are around 100fs. To measure temporal population dynamics the path length of the pump pulse can be controlled with a delay stage to introduce a well-defined time delay $\Delta t$ with respect to the probe laser. The pump beam is switched on and off with an acousto-optical modulator at a frequency of 2MHz before being superimposed with the probe beam on a beam splitter. A half-wave plate ($\lambda /2$) is used to control the relative polarization angle between pump and probe beam. The two co-propagating beams enter a laser scanning microscope, are focused on the sample with a NA=0.7 microscope objective, interact with the sample, and the pump beam is rejected with a pump filter before the probe is detected with a photodiode (detector). The difference in absorption of the probe between periods when the pump is turned on or off is measured with a lock-in amplifier as a function of the inter-pulse delay. We chose the lock-in phase such that an increase of probe absorption when the pump is on corresponds to a positive sign and a decrease to a negative sign. A transient absorption curve is measured by sweeping the arrival time $\Delta t$ between pump and probe pulses. A TA image is measured by raster scanning the laser over the sample and measuring a TA curve at each position. The inset in Figure~\ref{fig:experimental_setup}b) displays a projection of a TA image of a primary melanoma and the blue curve represents its average TA curve, which is dominated by transient gain processes as the signal is negative. 

    \subsection{Melanoma Biopsies and Sample Preparation}
    Samples were obtained from formalin-fixed, paraffin-embedded tissue blocks of 71 primary cutaneous melanoma tumors and of 18 cutaneous melanoma metastases. Five out of the 71 primary tumors did not have enough melanin for imaging, leaving 34 primary tumors that developed metastasis and 32 primary tumors that did not (with a follow-up time of at least 5 years).
    
    Two consecutive serial sections of $5\mu$m thickness were prepared from each block. The first section was stained with hematoxylin and eosin (H\&E) for conventional histopathological assessment, while the second was deparaffinized for PPM imaging by heating in an oven to remove bulk paraffin, followed by immersion in xylene and graded alcohols to remove residual paraffin. The melanoma tumor region was annotated on the H\&E-stained section by a board-certified dermatopathologist. The position information of the annotated tumor region was transferred to the unstained serial section by visual alignment of anatomical landmarks. Pigmented areas within the tumor region of the unstained section were used for PPM imaging.

    \subsection{Data Acquisition}
    PPM images for the 770/730 wavelength combination were acquired at 64 time delays spanning $-15$ to $80$\,ps with parallel pump-probe polarization. For 730/810, each region of interest (ROI) was imaged twice, once with parallel and once with perpendicular polarization, at 25 time delays spanning $-5$ to $110$\,ps.

    All acquisitions used an average power of 0.3\,mW per beam, measured after the objective with a beam waist of $w_0 \approx 0.5\,\mu$m, and a pixel dimension of 300\,nm, slightly undersampling with respect to the optical resolution. For each sample, several ROIs were imaged within the pathologist-annotated tumor region wherever melanin was present, with the field of view matched to the pigmented area. Because the amount of melanin differs between tumors, the number of ROIs varies substantially across the cohort, which motivates using patient weights during machine learning, as described in Section~\ref{ch:classification}.

    Pulse widths and three-dimensional overlap of pump and probe beams were characterized before every imaging session to ensure consistent temporal and spatial overlap of pump and probe beams across the full acquisition period. Average pulse widths were $(77 \pm 3)$\,fs for 770/730 and $(66 \pm 4)$\,fs for 730/810, beam overlap was characterized and optimized with point spread function measurements, a more detailed description can be found in the Supplementary Section \ref{si_ch:laser}.

    \subsection{Data Processing}
    Raw PPM stacks (sets of images at sthe chosen time delays) were processed as follows. Surgical ink used during surgery to mark excision boundaries occasionally remains in the unstained sections and produces an artificial signal. Therefore, all stacks were visually screened and ink-containing regions manually cropped (Supplementary Section \ref{si_ch:ink}). Excited-state dynamics with lifetimes much longer than the 12.5\,ns pulse train repetition time manifest as a constant offset, which was removed by subtracting the mean of the images acquired at negative time delays ($\Delta t\leq -1$\,ps), so that the resulting TA signal is zero at $\Delta t=0$  (see Figure~\ref{fig:experimental_setup}b). In each stack, regions with little or no melanin signal were excluded using an intensity mask at a heuristically chosen threshold that balances background rejection against retention of genuine melanin signal (Supplementary Section \ref{si_ch:background}).

    For 730/810 acquisitions, the parallel and perpendicular stacks were decomposed into ground state bleaching (GSB) and excited state absorption (ESA) components, which carry different polarization dependencies in melanin \cite{Grass2022b}:
    \begin{align}
        s_\text{GSB} &= \tfrac{3}{2}\,s_\text{para} - \tfrac{3}{2}\,s_\text{perp}, \\
        s_\text{ESA} &= -\tfrac{1}{2}\,s_\text{para} + \tfrac{3}{2}\,s_\text{perp}.
    \end{align}
    This decomposition turns the two measured bipolar curves into two unipolar component dynamics (Supplementary Figure \ref{si_fig:decomposition}) and is also used to compute the ESA-to-GSB ratio analyzed in Section~\ref{ch:logreg}.

    Finally, each masked stack was partitioned into non-overlapping squares of side length $l_\text{px}$ pixels and the TA curves within each square spatially averaged. This procedure improves signal-to-noise ratio at the cost of reduced resolution. The parameter $l_\text{px}$ is treated throughout as a free parameter: we use 16, 32, 64, 128, 256, and 512 pixels, corresponding to squares of width 4.9, 9.7, 19.4, 38.7, 77.4, and 154.8\,$\mu$m.

    \subsection{Model Fitting}
    To reduce each binned TA curve to a small set of physically interpretable parameters, we fit it to an analytical excited-state model comprising a superposition of individual pump-probe interactions\cite{Grass2022b}. Nonlinear interactions populating a real excited state, such as GSB and ESA, contribute an exponential decay with characteristic lifetime $\tau$ (\emph{delayed} processes). Nonlinear interactions involving virtual energy states, such as two-photon absorption or stimulated Raman scattering, are instantaneous on the timescale of the pulse widths and are modeled as a delta function (\emph{instantaneous} processes). All interactions are convolved with the measured instrument response function, resulting in $s_\text{delayed}(\Delta t, \tau)$, its long-lifetime limit $s_\infty(\Delta t) = s_\text{delayed}(\Delta t, \tau\rightarrow\infty)$, and $s_\text{inst}(\Delta t)$ and are explicitly listed in Supplementary Section \ref{si_ch:model}.

    For 770/730 acquisitions we fit
    \begin{align}
        s(\Delta t) = -a_1\, s_\text{delayed}(\Delta t, \tau_1)
        - a_2\, s_\text{delayed}(\Delta t, \tau_2)
        - a_3\, s_\text{inst}(\Delta t)
        - a_4\, s_\infty(\Delta t),
    \end{align}
    comprising two exponential decays $\{a_1, \tau_1\}$ and $\{a_2, \tau_2\}$, one instantaneous component $a_3$, and one long-lived component $a_4$. The instantaneous term accounts for a weak spurious positive signal, most likely two-photon absorption from xylene, surrounding cells, or the microscope slide; its inclusion is favored by the Akaike information criterion over a variant without it.

    For 730/810 acquisitions, the decomposed ESA and GSB stacks are each fit independently with the same form:
    \begin{align}
        s(\Delta t) = -a_1\, s_\text{delayed}(\Delta t, \tau_1)
        - a_2\, s_\text{delayed}(\Delta t, \tau_2)
        - a_3\, s_\infty(\Delta t).
    \end{align}

    Parameters were estimated by nonlinear least-squares fitting with a Levenberg-Marquardt algorithm, yielding 6 parameters ($a_1, a_2, a_3, a_4, \tau_1, \tau_2$) per TA curve for 770/730 and 10 parameters (5 per component) for 730/810. Each curve is thus compressed from 64 sampled time delays (770/730) or $2 \times 25$ delays (730/810) to a physically interpretable set of parameters that serves as the feature vector for classification.

    \subsection{Classification}
    \label{ch:classification}
    We trained common machine learning classifiers to distinguish primary tumors of patients who developed metastatic disease from those who did not, using the fit parameters derived in the previous section as input features. Training was done in python using scikit-learn \cite{scikit-learn} version 1.7.2. With only 71 patients available, the amount of training data is too small to train convolutional neural network (CNN) architectures, which are the current state of the art for multi-dimensional image classification but require much larger training sets.
    
    A further complication is that only pigmented regions of a tumor produce usable PPM signal, so the number of TA curves obtained per patient varies substantially across the cohort. We addressed this imbalance with three strategies that operate at different levels of data aggregation:
    \begin{enumerate}
        \item \textit{Patient averaged fits:} All TA curves from a given patient are spatially averaged into a single TA curve, which is then fit with the model described above. When imaged with 770/730 (730/810), each patient contributes a single feature vector of 6 (10) fit parameters with a well-defined metastatic label.
    
        \item \textit{Patient level pooling:} Each TA curve from a patient inherits that patient's label (0 = non-metastatic, 1 = metastatic), and a classifier is trained on individual ROIs. To prevent patients with many ROIs from dominating the loss, each ROI is assigned a sample weight of $1/N_p$, with $N_p$ the number of ROIs for that patient. A patient's final label is obtained by combining all ROI predictions: if more than a fraction $r$ of the ROIs are individually classified as metastatic, the patient is labeled metastatic. We heuristically set $r = 0.25$, reflecting the assumption that metastatic potential may be signaled by only a subset of the tumor. This threshold-based group accuracy is used both as the scoring metric during cross-validation and as the reported test accuracy, and is implemented via scikit-learn's metadata routing.
    
        \item \textit{Bagged patient level features:} This strategy is a simple version of multiple instance learning 
        \cite{dietterichSolvingMultipleInstance1997, carbonneauMultipleInstanceLearning2018}. All ROI fit parameters from a given patient are grouped into a bag, and per-patient bag statistics are computed as features. Specifically, we compute the minimum, maximum, and mean of the fit parameters. Training and evaluation then proceed as in strategy (1).
    \end{enumerate}
    
    We benchmarked six commonly used classifiers on all three strategies: $k$-nearest neighbors (KNN), histogram-based gradient boosting, logistic regression, Gaussian naive Bayes, random forest, and a support vector classifier (SVC). For each classifier we performed a grid search over its principal hyperparameters (e.g. tree depth and ensemble size for the tree-based models, kernel bandwidth and regularization strength for SVC, number of neighbors and distance metric for KNN). Hyperparameter selection was performed with three-fold cross-validation on the training set, using StratifiedKFold for strategies (1) and (3), where each patient contributes a single sample, and StratifiedGroupKFold for strategy (2) to ensure that all ROIs from a given patient fall in the same fold. Features were standardized inside each pipeline prior to fitting where appropriate. For each combination of classifier and strategy, we repeated the following procedure 20 times to estimate variability: patients were split into training (70\%) and test (30\%) sets, stratified by metastatic status so that class proportions were preserved; the model was trained and hyperparameter-tuned on the training patients only; and patient-level accuracy was evaluated on the held-out test patients. The reported performance for each classifier is the mean and standard deviation of test accuracy across all repetitions. For strategies (2) and (3), which retain ROI-level information, we repeated this benchmark for square side lengths of $l_\text{px} \in \{16, 32, 64, 128, 256, 512\}$ pixels to study how the spatial averaging trades off SNR against the number of samples available for training. 
    
    We stress that this benchmark is not intended to produce a production-ready diagnostic model as the number of classifiers and hyperparameter combinations explored, relative to the cohort size, carries a substantial risk of overfitting. Our aim is rather to characterize how much of the metastatic signal is accessible from PPM-derived melanin dynamics and to identify the principal methodological bottlenecks.
    
\section{Results}
    \subsection{770/730 Pump-Probe Microscopy of Melanoma Tumors}
    \label{ch:770730_images}
\begin{figure}[h!]
        \centering
        \includegraphics{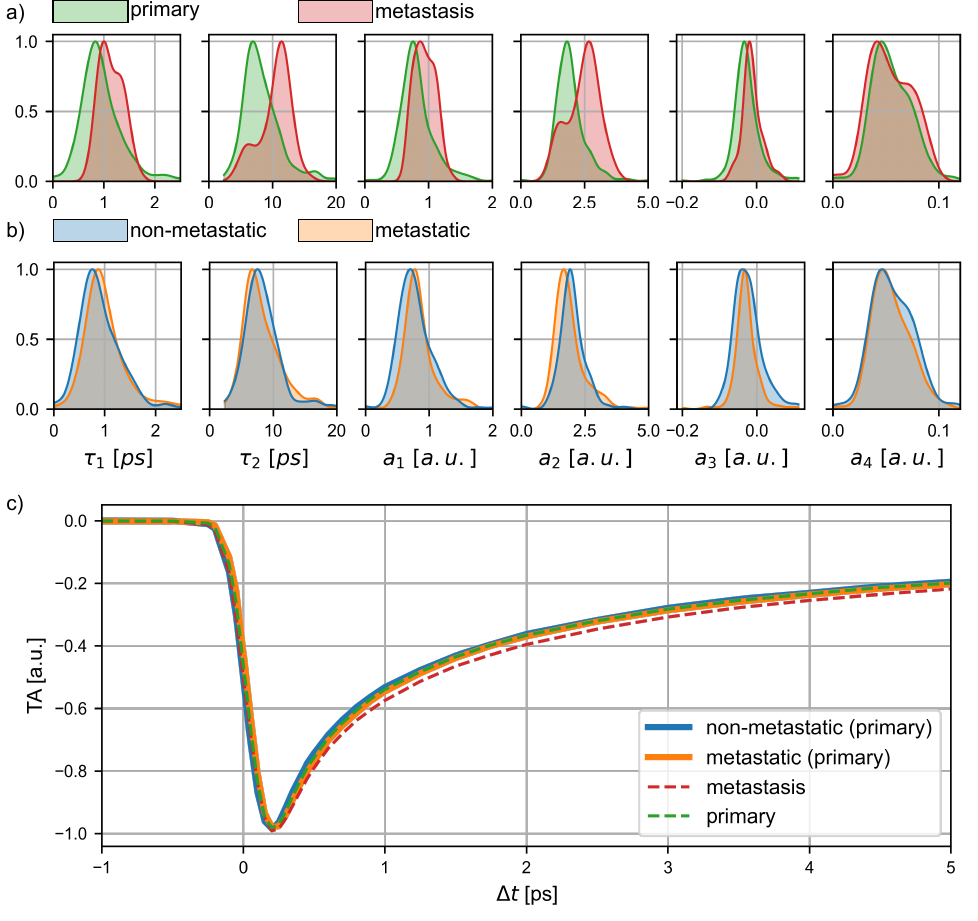}
        \caption{Kernel density estimates of the six fit parameters for ROIs with $l_\text{px} = 512$\,pixels: a) primary 
        tumors (green) versus metastases (red), and b) primary tumors from patients who did not develop metastatic disease (non-metastatic, blue) versus those who did (metastatic, orange). c) Average transient absorption curve of primary non-metastatic tumors (blue), primary tumors that developed metastases (orange), all primary tumors (green dashed) and metastasis (red dashed).}
        \label{fig:histo}
    \end{figure}

    Of the 71 primary tumor samples, 5 were amelanotic and did not produce a usable PPM signal, leaving 66 primary tumors and 18 metastatic tumors. Across these 84 samples, 411 ROIs were acquired and processed following the pipeline described in Section~\ref{ch:material_methods}. Here we compare the fit-parameter distributions between (i) primary and metastatic tumors and (ii) primary tumors from patients who subsequently developed metastases versus those who did not.
    Kernel density estimates of the six fit parameters ($\tau_1, \tau_2, a_1, a_2, a_3, a_4$) for a ROI side length of $l_\text{px} = 512$ pixels are shown in Figure~\ref{fig:histo}. The top row (green vs.\ red) compares primary tumors with metastases; the bottom row (blue vs.\ orange) compares primary tumors from patients who did and did not subsequently develop metastatic disease.

    Visually, the lifetimes $\tau_1$ and $\tau_2$ and their amplitudes $a_1$ and $a_2$ are distributed differently between primary and metastatic tumors: excited-state dynamics in primary tumors are generally faster and of lower amplitude than in metastases. In contrast, the distributions of the same parameters for primary tumors that did and did not subsequently metastasize are visually indistinguishable.
    
    To quantify these differences we computed the common-language effect size $f = U / (n_1 \cdot n_2)$, with $U$ the Mann-Whitney $U$ statistic and $n_1, n_2$ the sizes of the two groups. This quantity is equivalent to the area under the ROC curve and provides an intuitive interpretation: the probability that a randomly drawn value from group 1 exceeds a randomly drawn value from group 2. Values close to 0.5 indicate that the parameter cannot separate the two groups; values close to 0 or 1 indicate strong separation, in opposite directions. Table~\ref{tab:mann-whitney-u} lists $f$ for each fit parameter, together with $|f - 0.5|$ in parentheses, for both comparisons. The primary versus metastasis row is labeled ``location'' and the metastatic versus non-metastatic row is labeled ``metastatic''.
    \begin{table}[h]
        \centering
        \begin{tabular}{c|c|c|c|c|c|c|}
                    &  $\tau_1$    & $\tau_2$  & $a_1$ & $a_2$ & $a_3$ & $a_4$ \\
                \hline
            location    & 0.30 (0.20)  & 0.26 (0.24)  & 0.30 (0.20) & 0.27 (0.23) & 0.32 (0.18) & 0.52 (0.02) \\ 
            metastatic  & 0.41 (0.09)  & 0.53 (0.03)  & 0.41 (0.09) & 0.61 (0.11) & 0.51 (0.01) & 0.52 (0.03) \\
        \end{tabular}
        \caption{Common-language effect size $f = U/(n_1 n_2)$ for each 
        fit parameter, with $|f - 0.5|$ in parentheses. Top row: primary 
        tumors versus metastases (``location''); bottom row: primary 
        tumors from non-metastatic versus metastatic patients 
        (``metastatic'').}
        \label{tab:mann-whitney-u}
    \end{table}

    The effect sizes quantify the results displayed in Figure~\ref{fig:histo}. For the location contrast, $\tau_1, \tau_2$, $a_1$ and $a_2$ all deviate substantially from 0.5, indicating that melanin excited-state dynamics differ meaningfully between primary and metastatic lesions. Notably, the amplitude $a_4$ of the long-lived component is indistinguishable between the two populations ($f = 0.52$), suggesting that the distinguishing information lies in the fast excited-state dynamics rather than in the long-time offset. In contrast, the ``metastatic'' cohort yields effect sizes that mostly cluster near 0.5, with the largest deviation observed for $a_2$ ($f = 0.61$). Thus, on a single-parameter distributional level, PPM-derived melanin dynamics reliably distinguish\emph{primary from metastatic tissue}, but do not distinguish primary tumors by their eventual metastatic \emph{outcome}. This is also highlighted by figure \ref{fig:histo}c) showing that only the metastasis curve (red, dashed) does not significantly overlap with the remaining three curves of non-metastatic tumors, metastatic tumors, and primary tumors. This motivates the classification approach described in Chapter~\ref{ch:material_methods}, where combinations of parameters are used rather than any single feature.

    \subsection{730/810 Polarization-Resolved Pump-Probe Microscopy of Primary Melanoma Tumors}
    Polarization-resolved PPM was performed on a subset of 27 patients (15 non-metastatic primary tumors, 12 primary tumors that developed metastases), yielding 108 ROIs. For each ROI, two co-registered PPM stacks were acquired: one with parallel and one with perpendicular pump-probe polarization. The stacks were processed and decomposed into ESA and GSB components as described in Section~\ref{ch:material_methods}. In addition to the six per-component fit parameters, we computed a per-ROI ESA-to-GSB ratio, defined as the ratio of the area under the ESA TA curve to that under the GSB TA curve. Per-patient values were obtained by averaging over ROIs.
    
    \subsection{Univariate Logistic Regression}
    \label{ch:logreg}
    As a simple baseline, we tested whether individual fit parameters predict metastatic outcome using logistic regression at the patient level, prior to the classification benchmarks. Two models were fitted: one using the two excited-state lifetimes $\tau_1$ and $\tau_2$ from the 770/730 acquisitions (all 66 pigmented primary tumors), and one using the ESA-to-GSB ratio derived from the 730/810 polarization-resolved acquisitions (the 27-patient subset). 

    The lifetime model was not statistically significant relative to the null model, $\chi^2(2) = 3.73$, $p = 0.15$, and explained essentially none of the variance in outcome (McFadden's pseudo-$R^2 = 0.04$). Neither $\tau_1$ ($b = 0.57$, $\mathrm{SE} = 0.31$, $z = 1.83$, $p = 0.07$; OR $= 1.78$, 95\% CI $[0.96, 3.29]$) nor $\tau_2$ ($b = -0.3$, $\mathrm{SE} = 0.31$, $z = -0.98$, $p = 0.33$; OR $= 0.74$, 95\% CI $[0.40, 1.35]$) was significantly associated with metastatic outcome. The ESA-to-GSB-ratio model was likewise not statistically significant relative to the null model, $\chi^2(1) = 1.15$, $p = 0.28$, with McFadden's pseudo-$R^2 = 0.031$. The ratio was not significantly associated with metastatic outcome ($b = 0.65$, $\mathrm{SE} = 0.63$, $z = 1.05$, $p = 0.30$; OR $= 1.92$, 95\% CI $[0.56, 6.55]$).

    Taken together, these univariate analyses provide no evidence that the two excited-state lifetimes or the ESA-to-GSB ratio carry outcome information on their own. This is consistent with the distributional overlap observed in Figure~\ref{fig:histo} and motivates the multivariate classification approach in the following section, where the joint information across multiple fit parameters is exploited.

    \subsection{770/730 based Classifier}
    As described in Section~\ref{ch:material_methods}, we benchmarked the six classifiers with three data-aggregation strategies: patient-averaged fits, region-level classification with patient-level pooling (referred to as ``majority vote'' in the figure), and bagged patient-level features. For the two ROI-based strategies we swept the ROI side length over $l_\text{px} \in \{16, 32, 64, 128, 256, 512\}$~pixels; $l_\text{px} = 64$~pixels was the best-performing size and is shown in Figure~\ref{fig:770730_classifier}. Results for the remaining sizes are provided in the Supplementary Material \ref{si_ch:classifier}. Each classifier was retrained on 20 stratified 70/30 patient-level train--test splits; points and error bars in the figure show the mean and standard deviation of patient-level test accuracy across these repetitions. The dashed grey line marks chance-level performance.

    \begin{figure}[h]
        \centering
        \includegraphics{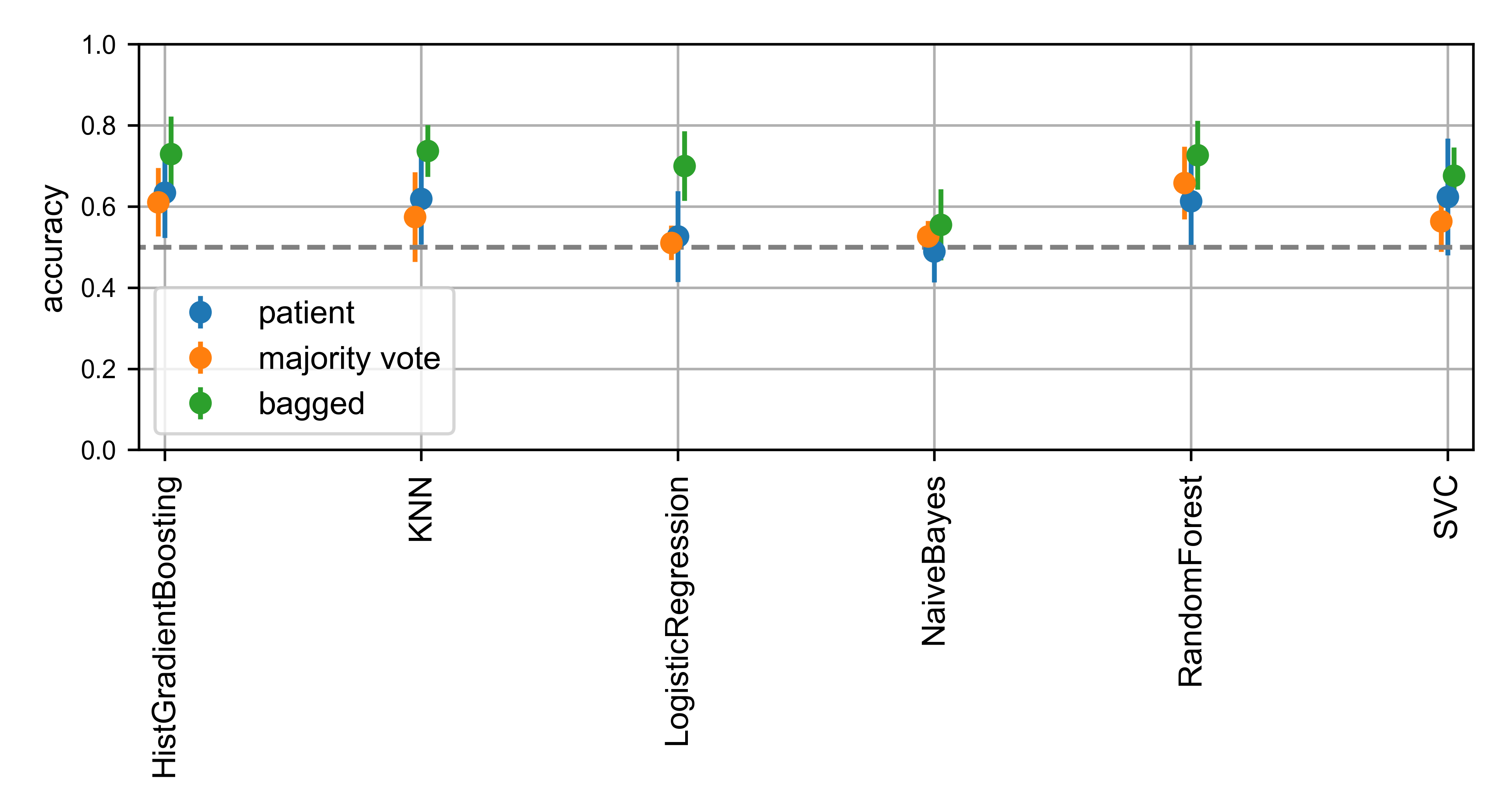}
        \caption{Patient-level test accuracy of the six classifiers under the three aggregation strategies on the 770/730 dataset: patient-averaged fits (blue), classification with patient-level threshold pooling (``majority vote'', orange), and bagged patient-level features (green). Points and error bars show the mean and standard deviation over 20 train--test split iterations; the dashed grey line indicates chance performance. Results shown for $l_\text{px} = 64$~pixels, other bin sizes are provided in the Supplementary Material \ref{si_ch:classifier}}
        \label{fig:770730_classifier}
    \end{figure}

    Several trends are apparent. First, the patient-averaged strategy (blue) performs poorly across all classifiers, with accuracies of 50--65\%, only marginally above chance. This is consistent with the distributional analysis in figure~\ref{fig:histo}, where individual fit parameters showed little separation between metastatic and non-metastatic primaries. Second, the majority-vote strategy performs generally poor with the exception of the Random Forest classifier reaching 70\% classification accuracy at a ROI size of 32. The bagged approach (green) reaches higher accuracies and performing best for every classifier tested. The best-performing model is KNN with 74\% accuracy followed closely by histogram gradient boosting and random forest at 73\%. Across the ROI-based strategies, accuracy improves as $l_\text{px}$ decreases from 512 to 64 pixels, peaks at 64 pixels, and degrades again for smaller sizes (see Supplementary Material \ref{si_ch:classifier}). This is consistent with a bias-variance trade-off: larger bins yield too few samples per patient for the pooling and bagging strategies to add information, while smaller bins average too few pixels to overcome noise in the fits. Notably, across very different model architectures and pooling schemes, performance appears to plateau near 74\% on held-out test patients, suggesting that this level reflects a ceiling imposed by cohort size and signal-to-noise limitations of the fitting procedure.

    Note that we trained all models at all sizes with and without parameter $a_3$, the fit amplitude of the instantaneous process parameterizing a weak spurious positive signal, most likely two-photon absorption from xylene, surrounding cells, or the microscope slide. Models including $a_3$ either perform worse or equally compared with models trained without $a_3$. All model performances are listed in detail in the supplementary information \ref{si_ch:classifier}.

    \subsection{730/810 Polarization-Resolved Classifier}
\begin{figure}[h]
        \centering
        \includegraphics{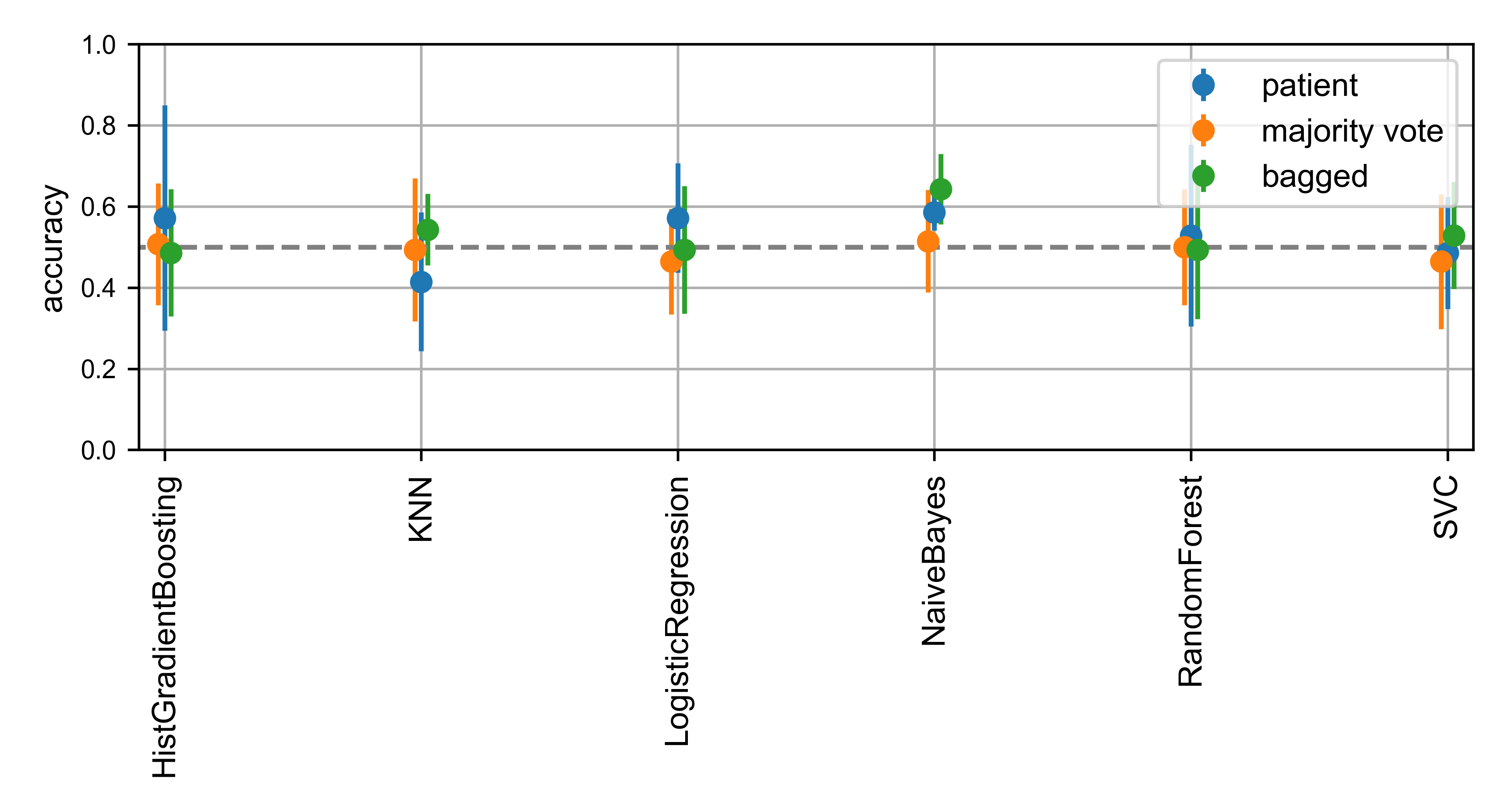}
        \caption{Patient-level test accuracy of the six classifiers on the 730/810 polarization-resolved dataset under the three aggregation strategies: patient-averaged fits (blue), classification with patient-level threshold pooling (``majority vote'', orange), and bagged patient-level features (green). Points and error bars show the mean and standard deviation over 20 train--test split iterations; the dashed grey line indicates chance performance. Results shown for the best-performing bin size of ($l_\text{px} = 128$~pixels), other bin sizes are provided in the Supplementary Material \ref{si_ch:classifier}}
        \label{fig:730810_classifier}
    \end{figure}

    For the polarization-resolved 730/810 acquisitions, we repeated the classification benchmark described above using the ESA and GSB fit parameters as input features. Because polarization-resolved data were acquired for only     27 patients, this analysis is more susceptible to sampling variability than the 770/730 benchmark on 66 patients. The main results are summarized in Figure~\ref{fig:730810_classifier}, which shows the best-performing ROI side length ($l_\text{px} = 128$~pixels).    
    
    In contrast to the 770/730 benchmark, no classifier--strategy combination achieves accuracy clearly above chance on the 730/810     polarization-resolved data. Most points cluster around 0.5, and the error bars for the patient-averaged strategy are very wide, reflecting the fact that a 30\% test split of 27 patients yields only about eight held-out patients per repetition. The one apparent outlier is Gaussian naive Bayes, which reaches approximately 0.64--0.65 under bagged strategies for the two largest ROI sizes; however, given that naive Bayes performed at chance level in the 770/730 benchmark (Figure~\ref{fig:770730_classifier}) and given the small cohort size, we interpret this result with caution rather than as evidence of a robust signal. Taken together, the 730/810 polarization-resolved data as currently available do not provide evidence that the ESA/GSB-decomposed excited-state dynamics carry information to identify metastatic melanoma.

\section{Discussion and Conclusion}

Identifying which primary cutaneous melanomas will progress to metastatic disease remains a central clinical challenge, since it determines which patients are candidates for immunotherapy or targeted therapy. In this manuscript we asked whether melanin excited-state dynamics measured by PPM in primary biopsies carry information about metastatic outcome, and we benchmarked several machine learning approaches to quantify that information.

The working hypothesis is that the aggressive tumor microenvironment subtly alters melanin structure, and that these alterations can be used to predict subsequent metastatic behavior. Specifically, the harsh intra- and extracellular environment is thought to disaggregate melanin from highly assembled protomolecules into smaller subunits \cite{Ju2018, Ju2019}. A large body of work on synthetic and natural melanin models \cite{simpsonNearInfraredExcitedState2014, Ju2018, Ju2019} indicates that intact, aggregated melanin exhibits faster excited-state decay which is consistent with its photoprotective role of rapidly converting absorbed light into heat, whereas disaggregated melanin generally show slower excited-state dynamics. From a biological standpoint, a melanocyte must accumulate many alterations before it can separate from the primary lesion, travel through lymphatic or blood vessels, and survive and replicate at a distant, non-physiological site. This picture presents two coupled challenges: identifying the molecular melanin features that predict metastasis, and localizing them within the primary tumor. The metastatic origin within a primary lesion may occupy only a small part of the full tumor volume, and the relevant molecular signature must additionally be distinguished from unrelated sources of melanin heterogeneity such as sun exposure, metal chelation, trauma, and melanin digested by melanophages, all of which modify melanin signals \cite{simpsonNearInfraredExcitedState2014, Grass2022b}. To avoid the localization problem, we began by imaging melanin in metastatic lesions themselves because these lesions should show metastatic features most clearly.

Consistent with this hypothesis, the distributional analysis in Figure~\ref{fig:histo} and the effect-size analysis in Section~\ref{ch:770730_images} show that excited-state dynamics of melanin in primary tumors differ substantially from those in metastases. In particular, both lifetimes are shorter in primaries than in metastases, consistent with the model in which primary tumor melanin is more aggregated and metastatic tumor melanin more disaggregated. However, the same features that separate primary from metastatic tissue at the cohort level fail to predict metastatic outcome at the individual patient level. The most straightforward interpretation is a needle-in-the-haystack effect: the metastatic-origin region within a primary tumor may be small relative to the total pigmented area, so its distinguishing features are diluted by the surrounding ``bulk'' primary tumor melanin or the imaged sections may not have contained the origin at all. Alternative explanations we cannot exclude are that the molecular signal predictive of metastasis is not acquired until the disaggregation process is well underway at the metastatic site, or that our fit-parameter representation does not capture the relevant features even when they are present. Distinguishing these possibilities will likely require volumetric imaging of the entire tumor and/or substantially larger patient cohorts.

This background clarifies why the simple univariate approaches suggested in earlier small-cohort work \cite{Ju2019, Grass2022b}, namely the excited-state lifetime for 770/730 imaging and the ESA-to-GSB ratio for 730/810 polarization-resolved imaging fail here. The logistic regression analyses in Section~\ref{ch:logreg} found no statistically significant association between either predictor and metastatic outcome in our cohort. Given the small sample sizes involved (66 and 27 patients respectively), we interpret this as evidence that these single features are insufficient rather than uninformative in an absolute sense, but the practical implication  is that univariate PPM-derived predictors are unlikely to be clinically useful on their own.

When applied a suite of common machine learning classifiers, using three data-aggregation strategies to accommodate the variable number of ROIs per patient, the best-performing classifiers reached approximately 74\% patient-level accuracy on a nearly balanced cohort (34 metastatic versus 32 non-metastatic), with the bagged approach performing best . Extensive hyperparameter tuning and comparison across model families gave similar peak performance, suggesting that 74\% represents a ceiling imposed by the current dataset rather than by model capacity. At the same time, we observed substantial variation in the optimal hyperparameters across train--test splits, which we interpret as indicating that the classifiers have not fully generalized and that a larger cohort would likely stabilize them and increase classification accuracy.

A further clinical application suggested by our findings is distinguishing primary from metastatic melanoma in diagnostically ambiguous biopsies. Some pigmented lesions cannot be confidently classified as primary or metastatic by histopathology alone, most often when a dermal-based melanocytic proliferation lacks a clear epidermal component, and such cases are sometimes reported as ``primary melanoma, metastasis cannot be excluded.'' This is not rare: melanoma of unknown primary accounts for roughly 13--18\% of patients presenting with regional or distant metastatic disease and is commonly attributed to complete regression at the primary site \cite{tarhiniImprovedPrognosisEvidence2022}, a mechanism that could plausibly produce exactly this kind of ambiguous lesion. The distinction matters clinically, separating a curable primary managed by excision from systemic metastatic disease eligible for adjuvant therapy. Our finding that melanin dynamics robustly separate primary from metastatic tissue at the cohort level (Section~\ref{ch:770730_images}) suggests PPM could in principle assist in this setting. 

Given current trends in medical image classification, the ideal architecture for this type of data would probably be a convolutional neural network with a multiple-instance-learning aggregation layer, which could exploit spatial structure that we discarded by binning and averaging. Training such a model from scratch is not feasible at our current cohort size. Models for skin cancer classification, trained on dermoscopic or clinical photographs, have been reported to achieve AUCs of 0.834--0.964 for metastatic melanoma prediction \cite{kulkarni_deep_2020, yan_deep_2025, lallas_deep_2024, lallas_deep_2025}, significantly higher than our best performing model with AUC=0.73 (corresponding to 74\% accuracy). These photograph-based models rely on tumor morphology, while our approach, in which we deliberately average away morphology, uses only molecular information about the melanin itself. We speculate that the two sources of information are complementary, and that a combined molecular-plus-morphological predictor could outperform either modality alone. Testing this directly would be a natural next step. Beyond outcome prediction, the robust separation of primary from metastatic melanin dynamics observed here raises the prospect of PPM as an ancillary tool for diagnostically indeterminate primary-versus-metastatic biopsies, a clinically consequential scenario our current cohort was not designed to test.

\subsection{Disclosures}
This manuscript was prepared using Overleaf including its built-in research writing toolkit that offers LaTeX code generation, error troubleshooting, and context-aware language editing directly inside the editor. Furthermore Claude Opus 4.8 chat interface was used for grammar and spell checking.
\subsection{Code, Data, and Materials Availability}
The imaging data supporting this study contain patient-derived specimen information and are therefore not publicly available due to privacy restrictions. De-identified transient absorption datasets, extracted fit parameters, and the analysis and classification code are available from the corresponding author upon reasonable request, subject to institutional data-use agreement and IRB approval.
\subsection{Funding}
This work was supported by the National Institutes of Health (NIH) under award 1R21-CA280272-01A1 and by the U.S. Department of Defense through the U.S. Army Medical Research Acquisition Activity (USAMRAA) under award ME220082.

\bibliography{report}   
\bibliographystyle{spiejour}   

\clearpage
\setcounter{section}{0}
\setcounter{figure}{0}
\setcounter{table}{0}
\setcounter{equation}{0}
\renewcommand{\thesection}{S\arabic{section}}
\renewcommand{\thesubsection}{S\arabic{section}.\arabic{subsection}}
\renewcommand{\thefigure}{S\arabic{figure}}
\renewcommand{\thetable}{S\arabic{table}}
\renewcommand{\theequation}{S\arabic{equation}}

\phantomsection
\addcontentsline{toc}{section}{Supplementary Information}
\begin{center}
  {\LARGE\bfseries Supplementary Information\par}
  \vspace{0.6em}
  {\large Evaluating the predictive power of pump-probe imaging contrast\\
   of melanin for metastatic outcome\par}
  \vspace{0.4em}
\end{center}
\vspace{1.5em}
\section{Laser Characterization}
\label{si_ch:laser}
The temporal and spatial characteristics of the pump and probe beams were measured before imaging sessions to ensure consistent data quality across the full acquisition period, which spanned several years.

\subsection{Pulse widths}
Pulse widths were determined from a TA curve of a Rhodamine 6G solution, whose transient absorption at these wavelengths is dominated by two-photon absorption. Because this process is instantaneous relative to the pulse duration, the measured TA curve approximates the intensity cross-correlation of the pump and probe pulses, from which the individual pulse widths are estimated assuming Gaussian pulse envelopes.

\begin{figure}[h!]
    \centering
    \includegraphics{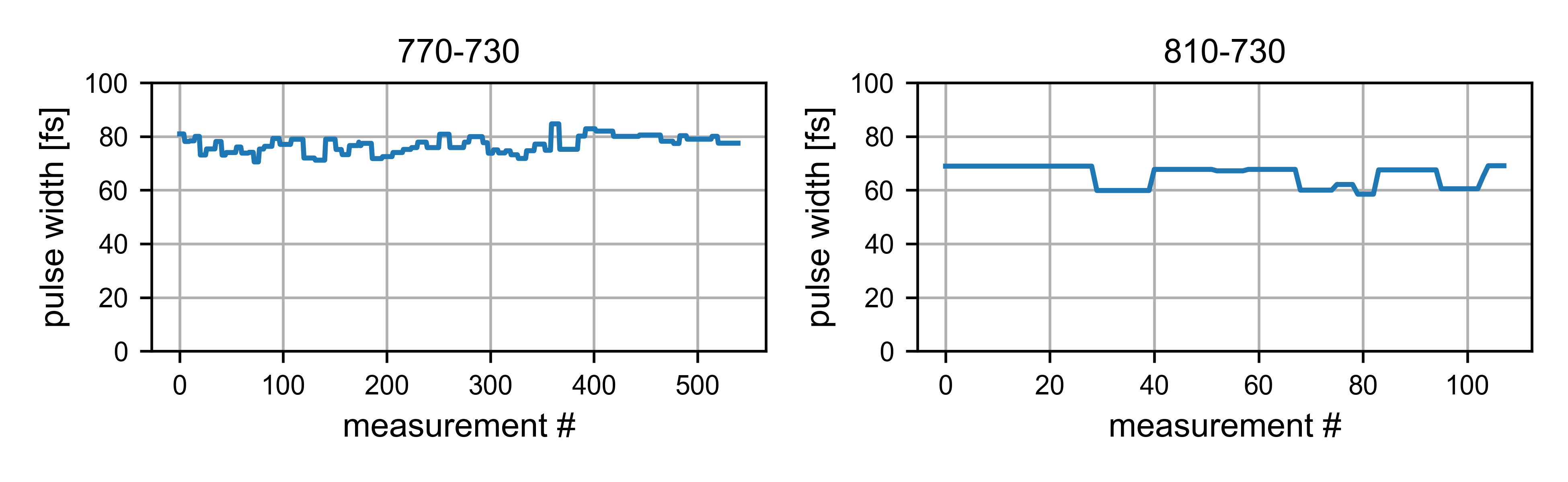}
    \caption{Cross-correlation measurements across the acquisition period. Pulse widths were extracted from the two-photon absorption response of Rhodamine 6G, assuming Gaussian pulse envelopes, for the 770/730 and 730/810 wavelength combinations.}
    \label{fig:supp_xcorr}
\end{figure}

For $\lambda_\text{pump} = 770$\,nm and $\lambda_\text{probe} = 730$\,nm the average pulse width was $\bar{t}_\text{pulse} = (77 \pm 3)$\,fs, with a minimum of 71\,fs and a maximum of 85\,fs. For $\lambda_\text{pump} = 730$\,nm and $\lambda_\text{probe} = 810$\,nm the average pulse width was $\bar{t}_\text{pulse} = (66 \pm 4)$\,fs, with a minimum of 58\,fs and a maximum of 69\,fs. Per-session measurements are shown in Figure~\ref{fig:supp_xcorr}. 

\subsection{Spatial overlap.}
To verify spatial overlap, the three-dimensional focus positions of the pump and probe beams were independently localized by imaging a 500\,nm-diameter polystyrene bead via two-photon fluorescence. Beam alignment was iteratively
adjusted until the two focal positions agreed to within half a standard deviation of the localization uncertainty. Note that in an ideal point-spread function measurement a sub-diffraction limit sized bead is imaged. Here, however, we only want to co-localize the two beams which can be achieved with larger beads, too.

\subsection{Polarization control.}
For 770/730 acquisitions, the half-wave plate ($\lambda/2$) was aligned so that both beams were parallel polarized, and a polarizer (not shown in the setup sketch of the main text) was inserted before the microscope objective to further ensure parallel polarization. For 730/810 acquisitions, each ROI was imaged twice: the half-wave plate was first set for parallel polarization and then rotated so that pump and probe were perpendicular. No polarizer was used in this configuration.

\section{Surgical Ink Removal}
\label{si_ch:ink}
\begin{figure}[h]
    \centering
    \includegraphics{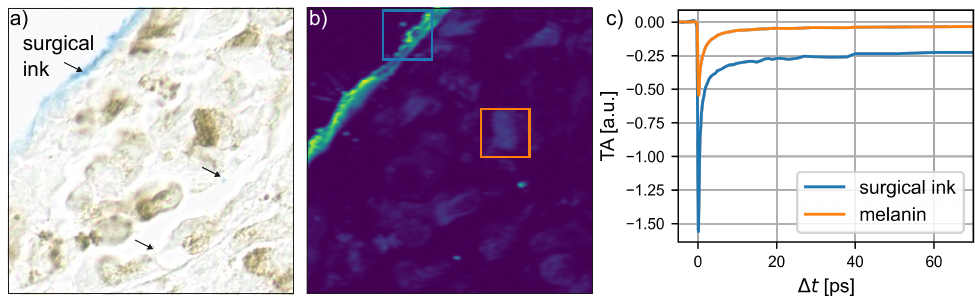}
    \caption{a) Unstained tumor section with residual blue surgical ink (black arrows) and melanin. b) Intensity projection of the PPM image stack of the adjacent unstained section; the ink produces a much stronger signal than melanin. c) Average TA curves from the ink-dominated (blue) and melanin-dominated (orange) regions boxed in b).}
    \label{fig:surgical_ink}
\end{figure}

Surgical ink used to mark tumor boundaries during excision occasionally persists in the unstained sections and produces a strong pump-probe signal unrelated to melanin, which must be removed prior to analysis. An example is shown in Figure~\ref{fig:surgical_ink}. In the unstained section (a), the blue-tinted regions indicated by black arrows correspond to residual ink, while the brown areas are melanin. In the intensity projection of the corresponding PPM stack (b), the ink signal substantially exceeds that of melanin in magnitude. The blue and orange boxes in (b) mark ink- and melanin-dominated regions respectively; their average TA curves (c) show clearly distinct dynamics. All stacks were visually screened and regions containing surgical ink were manually cropped.

\section{Baseline Subtraction and Intensity-Based Masking}
\label{si_ch:background}
Excited-state dynamics with lifetimes much longer than the 12.5\,ns inter-pulse spacing of the 80\,MHz laser manifest as a constant offset in the TA signal. This offset was removed by averaging the images acquired at negative time delays ($\Delta t \leq -1$\,ps) and subtracting the result from the entire stack. This corresponds to the first five images for 770/730 acquisitions and the first four for 730/810 acquisitions. 

To exclude regions with little or no melanin signal, an intensity mask was constructed for each stack by summing the absolute signal across all time delays, applying a Gaussian smoothing filter with a kernel size of 2 pixels, and thresholding at a value of 2 (in arbitrary units). This threshold was chosen empirically to balance rejection of background regions against retention of genuine melanin signal.

\section{Excited-State Model: Functional Forms}
\label{si_ch:model}
The main text describes the excited-state model conceptually. The explicit functional forms are provided here. Each contribution to the TA signal is convolved with the instrument response function of the microscope, which is itself the convolution of the pump and probe temporal envelopes. Assuming Gaussian envelopes, this double convolution is equivalent to a single convolution with a Gaussian of width $t_p = \sqrt{t_\text{pump}^2 + t_\text{probe}^2}$.

For a delayed process with lifetime $\tau$, the resulting TA signal as a function of time delay $\Delta t$ is
\begin{align}
    s_\text{delayed}(\Delta t, \tau) = \tfrac{1}{2}\,
    e^{-\Delta t/\tau + t_p^2/(2\tau^2)}
    \left[ 1 + \erf\!\left(
        \frac{\Delta t}{\sqrt{2}\, t_p} - \frac{t_p}{\sqrt{2}\, \tau}
    \right) \right].
\end{align}
When the lifetime is much longer than the acquisition window, the signal is approximated by the limit $\tau \to \infty$,
\begin{align}
    s_\infty(\Delta t) = \tfrac{1}{2}\!\left[ 1 +
    \erf\!\left( \frac{\Delta t}{\sqrt{2}\, t_p} \right) \right],
\end{align}
and for an instantaneous process the signal is Gaussian, 
\begin{align}
    s_\text{inst}(\Delta t) \propto \frac{1}{\sqrt{2\pi\, t_p^2}}\,
    e^{-\Delta t^2/(2t_p^2)}.
\end{align}

\section{Polarization Decomposition}
\label{si:decomposition}
\begin{figure}[h!!!]
    \centering
    \includegraphics{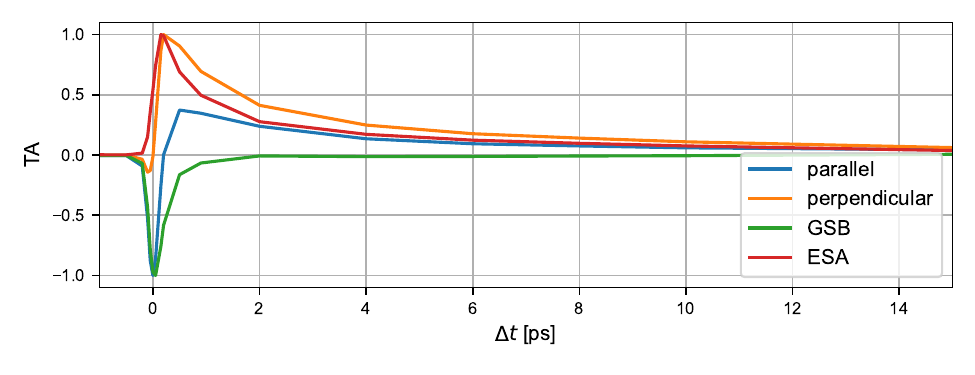}
    \caption{Polarization-based decomposition of pump-probe signals into GSB and ESA components. The bipolar TA curves measured with parallel (blue) and perpendicular (orange) pump-probe polarization are linearly combined to yield the unipolar GSB (green) and ESA (red) dynamics.}
    \label{si_fig:decomposition}
\end{figure}

Figure~\ref{si_fig:decomposition} illustrates the polarization-based decomposition applied to the 730/810 acquisitions. The bipolar TA curves measured with parallel and perpendicular pump-probe polarization are linearly
combined, using the relations given in the main text, into the two unipolar GSB and ESA components.

\newpage
\section{Classifier Performance Across Bin Sizes}
\label{si_ch:classifier}
\begin{figure}[h]
    \centering
    \includegraphics{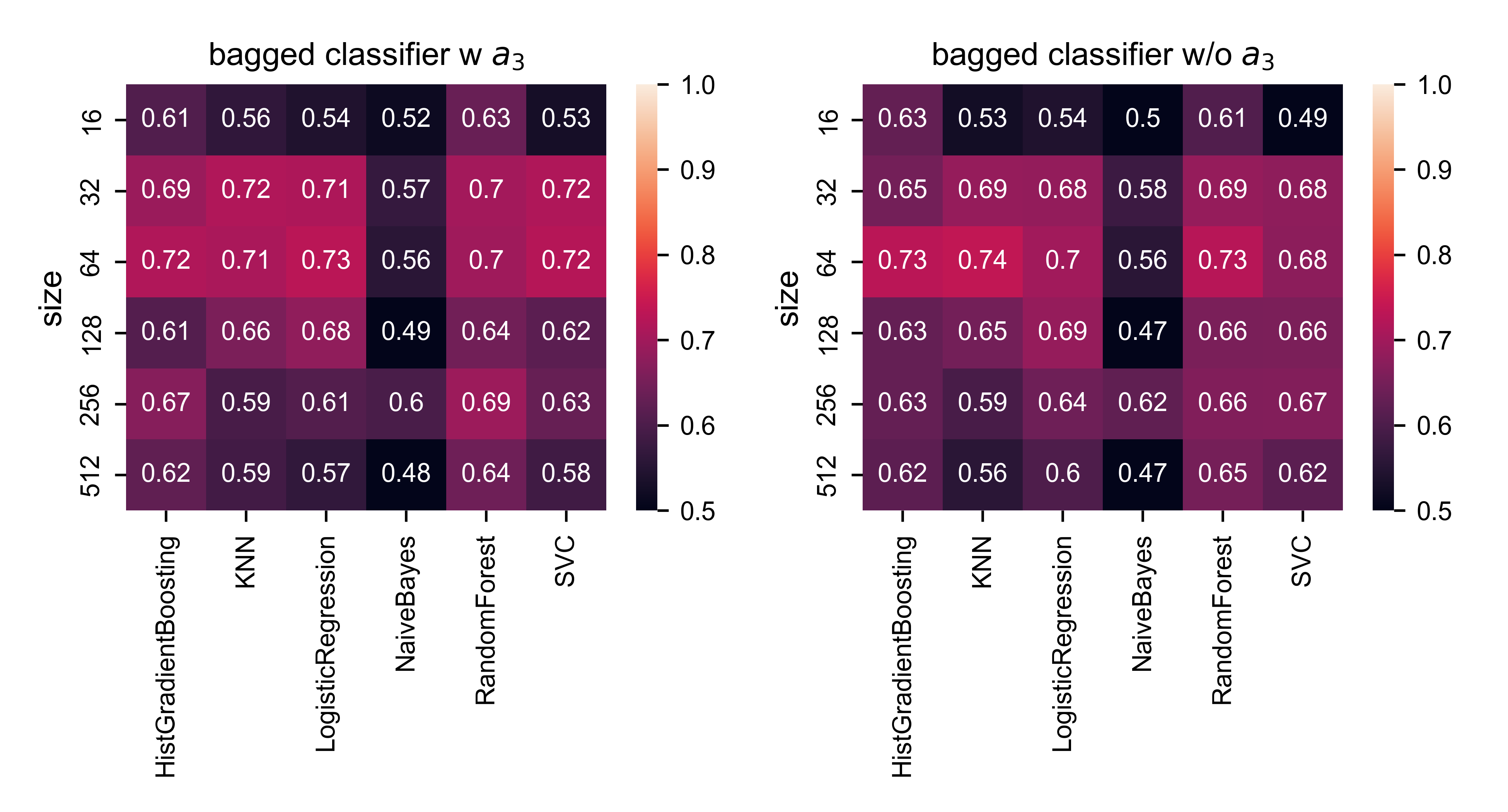}
    \caption{Patient-level test accuracy of the six classifiers on the 770/730 dataset for all bin sizes $l_\text{px}$ under bagging strategies including $a_3$ (labeled "w $a_3$") and excluding $a_3$ (labeled "w/o $a_3$")}
    \label{fig:770730_bagged}
\end{figure}

\begin{figure}[h]
    \centering
    \includegraphics{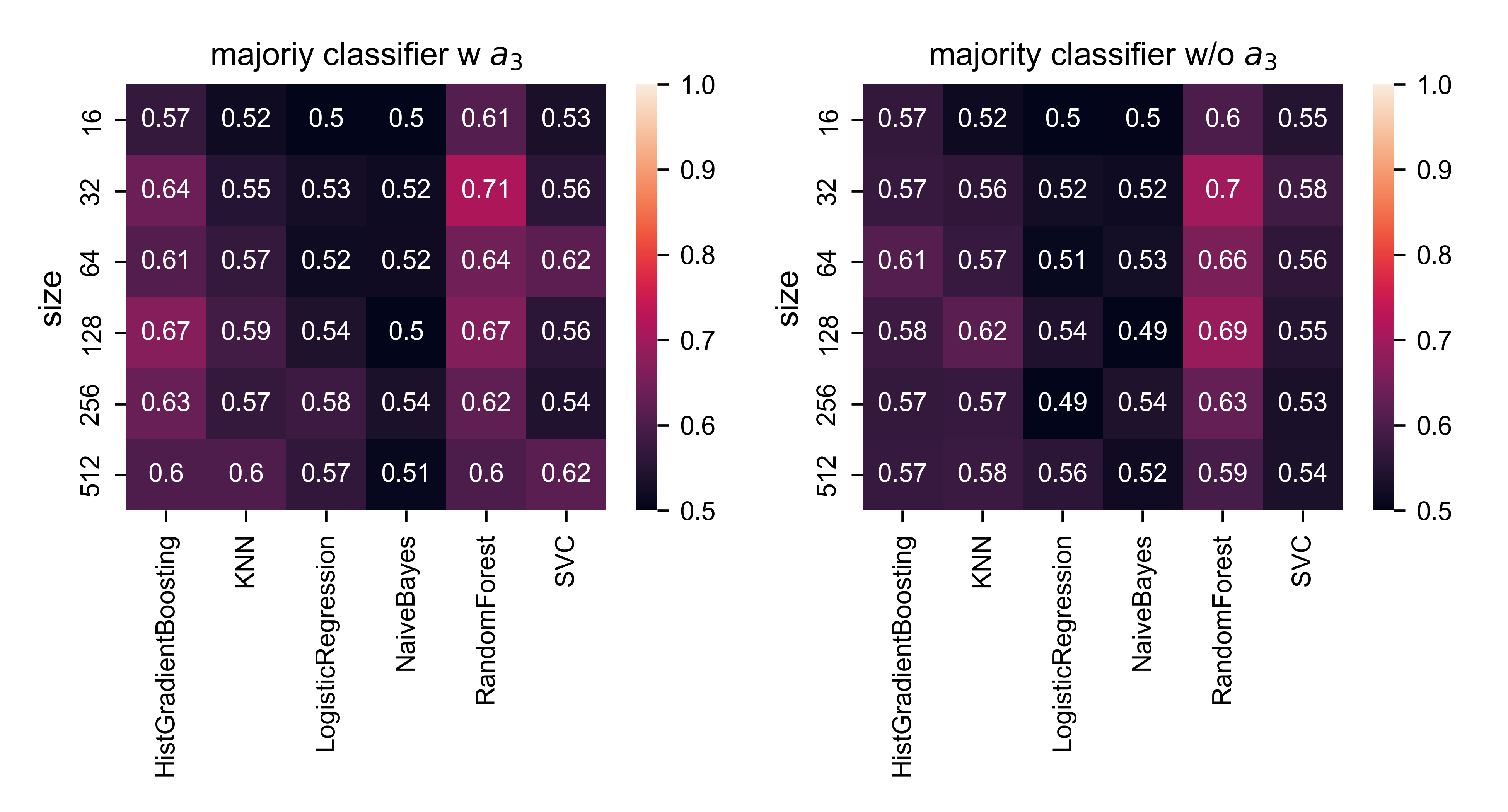}
    \caption{Patient-level test accuracy of the six classifiers on the 770/730 dataset for all bin sizes $l_\text{px}$ under majority voting strategies including $a_3$ (labeled "w $a_3$") and excluding $a_3$ (labeled "w/o $a_3$")}
    \label{fig:770730_vote}
\end{figure}

\begin{figure}[h]
    \centering
    \includegraphics{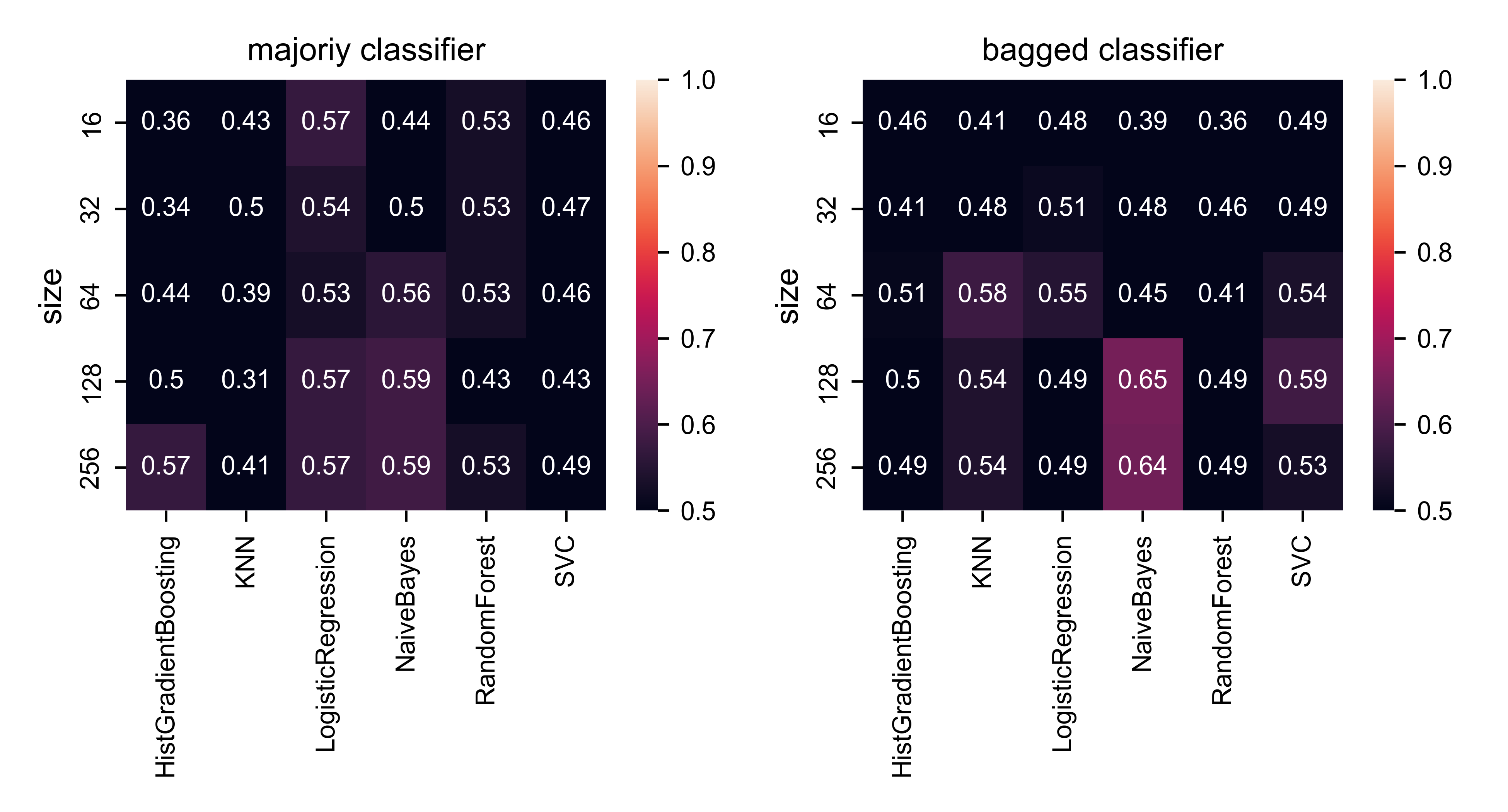}
    \caption{Patient-level test accuracy of the six classifiers on the 730/810 dataset for all bin sizes $l_\text{px}$ under majority voting strategy (left) and bagging strategy (right)}
    \label{fig:730810}
\end{figure}

The main text reports classifier performance at the best-performing bin size for each wavelength combination. Figures~\ref{fig:770730_bagged}, \ref{fig:770730_vote} and \ref{fig:730810} show the full sweep over $l_\text{px} \in \{16, 32, 64, 128, 256, 512\}$ pixels. 

\end{document}